\documentclass[aps,prb,twocolumn,eqsecnum,floatfix,superscriptaddress]{revtex4}
\usepackage{amssymb}
\usepackage{tabularx}
\usepackage{graphicx}
\usepackage{bm}
\usepackage{amsfonts}
\usepackage{amsmath}
\usepackage{exscale}
\usepackage{amsbsy}

\def\be{\begin{equation}}
\def\ee{\end{equation}}
\def\ba{\begin{eqnarray}}
\def\ea{\end{eqnarray}}

\def\C60{A$_x$C$_{60}$}

\def\HgCu3{HgCa$_2$Cu$_3$O$_{8+y}$}
\def\HgCu4{HgBa$_2$Ca$_3$Cu$_4$O$_{10+y}$}
\def\TlCu{Tl$_2$Ba$_2$CuO$_{6+\delta}$}
\def\TlCu3{Tl$_2$Ba$_2$Ca$_2$Cu$_3$O$_{10+y}$}
\def\TlCu4{Tl$_2$Ba$_2$Ca$_3$Cu$_4$O$_{12+y}$}

\def\BiCu3{Bi$_2$Sr$_2$Ca$_{2}$Cu$_3$O$_y$}

\def\8LSCO{La$_{1.88}$Sr$_{.12}$CuO$_4$}
\def\110LNSCO{La$_{1.5}$Nd$_{0.4}$Sr$_{0.1}$CuO$_{4}$}
\def\stage4LCO{La$_{2}$CuO$_{4+\delta}$}
\def\Y248{YBa$_2$Cu$_4$O$_8$}

\def\NbSe2{NbSe$_2$}
\def\TaSe2{TaSe$_2$}
\def\TiSe2{TiSe$_2$}
\def\NaCoOH2O{Na$_{0.3}$CoO$_{2y}$H$_2$O}
\def\MgB2{MgB${}_2$}

\def\s12{spin-${\frac{1}{2}}$}

\begin{document}

\title{Theory of the striped superconductor}
\author{ Erez Berg}
\affiliation{Department of Physics, Stanford University, Stanford, California 94305-4060}
\author{ Eduardo Fradkin}
\affiliation{Department of Physics, University of Illinois at Urbana-Champaign, Urbana,
Illinois 61801-3080}
\author{ Steven A. Kivelson}
\affiliation{Department of Physics, Stanford University, Stanford, California 94305-4060}
\date{\today }

\begin{abstract}
We define a distinct phase of matter, a \emph{pair density wave} (PDW), in
which the superconducting order parameter, $\phi(\vec r,\vec r^{\,\prime})$,
varies periodically as a function of position such that when averaged over
the center of mass position, $(\vec r + \vec r^{\,\prime})/2$, all
components of $\phi$ vanish identically. Specifically, we study the
simplest, unidirectional PDW, the ``striped superconductor,'' which we argue
may be at the heart of a number of spectacular experimental anomalies that
have been observed in the failed high temperature superconductor, {La$_{2-x}$%
Ba$_x$CuO$_4$}. 
We present a solvable microscopic model with strong electron-electron
interactions which supports a {PDW} groundstate. We also discuss, at the
level of Landau theory, the nature of the coupling between the {PDW} and
other order parameters, and the origins and some consequences of the unusual
sensitivity of this state to quenched disorder.
\end{abstract}

\maketitle



\section{A new phase of matter}

Superconductivity which arises from the pairing of electrons in time
reversed states is well understood as a weak coupling ``Fermi surface''
instability -- a consequence of arbitrarily weak effective attractive
interactions.\cite{Schrieffer64} In this paper we explore a new type of
superconducting state, a ``pair density wave'' (PDW), which is a distinct
state of matter, and which does not occur under generic circumstances in the
weak coupling limit -- it requires interactions in excess of a critical
strength. The {PDW} spontaneously breaks the global gauge symmetry, in
precisely the same way as a conventional superconductor does. Thus, the
order parameter is a charge 2$e $ complex scalar field, $\phi$. However, it
also spontaneously breaks some of the translational and point group
symmetries of the host crystal, in the same way as a conventional
charge-density wave (CDW). In a PDW, for fixed $\vec r -\vec r^{\,\prime}$,
the order parameter
\begin{equation}
\Delta(\vec r,\vec r^{\,\prime}) \equiv \langle \psi_{\uparrow}^\dagger(\vec
r)\psi_{\downarrow}^\dagger(\vec r^{\,\prime})\rangle  \label{eq:op-delta}
\end{equation}
is a periodic function of the center of mass coordinate, $\vec R \equiv(\vec
r + \vec r^{\,\prime})/2$. Here, $\psi_{\sigma}^\dagger(\vec r)$ is the
electron creation operator at position $\vec r$ with spin polarization $%
\sigma$. We will focus on the case in which translational symmetry is broken
only in one direction, \textit{i.e.} a unidirectional {PDW} or equivalently
a ``striped superconductor.''

Accompanying the {PDW} there is, as we will show below, induced CDW order
with half the period of the PDW. However, the {PDW} differs from a state of
coexisting superconducting and CDW order\cite{PDW}, which has also been
previously called a ``pair-density wave''.\cite{foot1} Unlike the
pair-density-wave state of Ref.[\onlinecite{PDW}], the average value of the
superconducting order parameter vanishes for the PDW state we discuss here.
This is a defining symmetry property of this state. The {PDW} is more
closely analogous to the Fulde-Ferrell-Larkin-Ovchinnikov (FFLO) state \cite%
{FFLO} which arises, under appropriate circumstances, when the electron gas
is partially polarized by an applied magnetic field. However, explicit
time-reversal symmetry breaking is an essential ingredient of the FFLO
state, and is responsible for lifting the degeneracy (``nesting'') between
time-reversed pairs of quasiparticle states. In the absence of quenched
disorder, the {PDW} state preserves time-reversal symmetry.\cite{foot2}

In general, there is no necessary relation between spin-density wave (SDW)
and {PDW} order -- whatever relation there is derives from common
microscopic physics rather than from general symmetry conditions. However,
the two orders can be linked if one postulates a larger, emergent SO(5)
symmetry\cite{zhangscience} which unifies the superconducting and SDW
orders. In this context, an early speculation\cite{zhangspiral} concerning
the existence of a unidirectional SO(5) spiral density wave state is
particularly interesting.\cite{foot3} Such a state would consist of
interleaving SDW and superconducting spirals, both with the same period. The
fact that this state is a spiral implies that the phase of $\phi $ varies as
a function of position, and hence that there are equilibrium currents. Thus,
the SO(5) spiral, though similar to a state of coexisting {PDW} and
collinear SDW order, is not the same. We have not found any microscopic
model that supports a spiral-{PDW} phase - whether coexisting with SDW
order, or not. However, it is an interesting state and deserves further
study.\cite{foot4} Many features of the superconducting state depend only on
the existence of a charge $2e$ order parameter. This includes zero
resistance, the Meissner effect, flux quantization and, in 3D, the existence
of a finite critical current. Therefore, these properties are expected in a {%
PDW} (in the absence of quenched disorder).

However, other properties are very different. For instance, in an s-wave
superconductor, there is a full gap in the quasiparticle spectrum. For a
d-wave superconductor, there are typically nodal points on the Fermi surface
which support gapless quasiparticle excitations, but the density of states
still vanishes at the Fermi energy, even in the presence of coexisting CDW
or most forms of SDW order\cite{bergandchen}. In contrast, in a {PDW}
portions of the Fermi surface are typically ungapped, which as in the case
of many CDWs, results in the reconstruction, but incomplete gapping of the
underlying Fermi surface\cite{bergandchen,dror}. In a conventional
superconductor, the anisotropy of the superconducting coherence length is
determined by the anisotropy of the single-electron effective mass; in a {PDW%
}, depending on the precise character of the translation symmetry breaking,
the anisotropy of the superconducting state can be parametrically larger
than the normal state resistivity anisotropy.

Perhaps the most significant new feature of a {PDW} superconductor is its
anomalous sensitivity to (non-magnetic) quenched disorder, in stark
comparison to a conventional superconductor. Specifically, although the
disorder potential does not couple directly to the superconducting order
parameter, it produces random variations in the magnitude \textit{and sign}
of the local superfluid density. Hence quenched disorder can readily drive a
{PDW} into a superconducting XY glass phase. (A similar effect was recently
predicted \cite{spivakoretoandme} in a d-wave superconductor near the
critical point of the superconductor to metal transition.)

Table \ref{tab:phases} characterizes the PDW, CDW, and uniform
superconducting (SC) phases by their order parameters and their
sensitivity to quenched disorder.  Here, $\Delta _{0}$ and $\Delta
_{Q}$ are, respectively, the uniform SC and PDW (modulated SC)
order parameters (where $Q$ is the ordering wavevector), and
$\rho_{Q}$ and $\rho_{2Q}$ are the two relevant CDW order
parameters. By definition, the pure PDW phase has non-zero $\Delta
_{Q}$ and a vanishing $\Delta _{0}$. It is thus different from a
phase of coexisting SC and CDW\ order, which has a
non-zero $\Delta _{0}$ \emph{and }$\Delta _{Q}$ (as well as a subsidiary $%
\rho _{Q}$ order). (In various places in Refs. [\onlinecite{PDW}],
as an aid to intuition, the SC+CDW and/or the CDW states have also
been referred to as \textquotedblleft Cooper pair
crystals\textquotedblright\.) As mentioned above, the main new
characteristic of the PDW (as opposed to a uniform SC) is its
sensitivity to disorder. Even weak disorder destroys the long
range PDW\ order, turning it into a glassy state. The properties
of these phases and their inter-relations will be derived from an
order parameter Landau theory in Section \ref{op}.

In a recent publication\cite{berg07} we argued that the remarkable
properties of {La$_{2-x}$Ba$_x$CuO$_4$} near $x=1/8$ filling can be
explained as a consequence of the symmetries of a striped superconducting
state. The main purpose of the present paper is to develop the theory of
this state.

This paper is organized as follows. In Section \ref{experiment} we give an
overview of the physics the striped superconductor {La$_{2-x}$Ba$_x$CuO$_4$}
evidenced in recent experiments\cite{tranquada07}. In Section \ref%
{microscopic} we discuss microscopic mechanisms that give rise to a PDW. In
Section \ref{models} we present several solvable models that shed light on
the physical origin of a {PDW} superconducting state. Here we investigate
the role of different model junctions separating two superconducting
regions, including an empty barrier (a trivial insulator) and an
antiferromagnetic insulating barrier. 
In both cases, under certain conditions, we find that a $\pi$ phase shift
between phases of the two superconductors is favored.
We then build on these insights to construct a model with a striped
superconducting ground state. We discuss qualitatively the quasiparticle
spectrum of a PDW state. In Section \ref{op} we present a Landau theory of
this state and discuss in detail the symmetry-dictated couplings between the
PDW order parameter and the nematic, charge density wave (CDW) and spin
density wave (SDW) order parameters characteristic of a stripe state. In a
recent publication, Agterberg and Tsunetsugu\cite{agterberg08} discussed a
Landau theory for a pair density wave which complements the results
presented in this Section. In Section \ref{glass_model} we present a
statistical mechanical model that describes a layer decoupled striped
superconductor, which is proposed to describe the phenomenology of {La$%
_{2-x} $Ba$_x$CuO$_4$}. In Section \ref{final} we discuss the implications
of this theory. Further details of the Landau theory are presented in
Appendix \ref{app:appendixA}. The case of coexisting uniform and striped
d-wave superconductivity is discussed Appendix \ref{app:uniform}.


\begin{table}[tbp]
\caption{Summary of the different phases discussed in the text and their
order parameters, as well as their sensitivity to quenched non-magnetic
disorder. A $\checkmark $ specifies that in a particular phase, the
corresponding order parameter is non-zero. See text for  details.}%
\begin{tabular}{|c||c|c|c|c||c|}
\hline
& $\Delta _{0}$ & $\Delta _{Q}$ & \multicolumn{1}{|c|}{$\rho _{Q}$} &
\multicolumn{1}{|c||}{$\rho _{2Q}$} & Sensitivity to disorder \\ \hline
PDW & $0$ & $\checkmark $ & $0$ & $\checkmark $ & Fragile \\ \hline
CDW & $0$ & $0$ & $\checkmark $ & $\checkmark $ & Fragile \\ \hline
CDW' & $0$ & $0$ & $0$ & $\checkmark $ & Fragile \\ \hline
SC & $\checkmark $ & $0$ & $0$ & $0$ & Robust \\ \hline
SC+CDW & $\checkmark $ & $\checkmark $ & $\checkmark $ & $\checkmark $ &
Becomes equivalent to SC \\ \hline
\end{tabular}%
\label{tab:phases}
\end{table}

\section{Striped superconductor in La$_{2-x}$Ba$_x$CuO$_4$}

\label{experiment}

La$_{2-x}$Ba$_x$CuO$_4$ is currently the most promising candidate
experimental system as a realization of a striped superconductor\cite%
{tranquada07,berg07}. Considerable indirect evidence in favor of the
existence of PDW order in this material has recently been gathered and
compiled in Ref. [\onlinecite{tranquada08}]. The putative PDW order has the
same period as the unidirectional (spin-stripe) SDW order which is known,
from neutron scattering studies, to exist in this material\cite%
{tranquada95,kivelson03}. Strong evidence that a striped SC can be induced
in underdoped {La$_{2-x}$Sr$_x$CuO$_4$} by the application of a transverse
magnetic field, which is also known to induce spin stripe order, has also
recently been presented in Ref. [\onlinecite{basov08}].

The behavior of {La$_{2-x}$Ba$_{x}$CuO$_{4}$} is very striking, and rather
complex. We will not elaborate upon it here (see Ref. [%
\onlinecite{tranquada08}]). However, there are two qualitative features of
the data on which we would like to focus: 1) With the onset of stripe spin
order at 42 K,\cite{foot5} there is a large (in magnitude) and strongly
temperature dependent enhancement of the anisotropy of the resistivity and
other properties, such that below 42 K the in-plane charge dynamics
resembles those of a superconductor, while in the c-direction the system
remains poorly metallic. The most extreme illustration of this occurs in the
temperature range $10\; \text{K} < T <16 \; \text{K}$, in which the in-plane
direction resistivity is immeasurably small, while the c-axis resistivity is
in the 1-10 m$\Omega $ range, so the resistivity anisotropy ratio is
consistent with infinity. 2) Despite the fact that many clear signatures of
superconductivity onset at temperatures in excess of 40 K, and that angle
resolved photoemission has inferred a ``gap'' \cite{valla06,shen08} of order
20 meV, the fully superconducting state (\textit{i.e.} the Meissner effect
and zero resistance in all directions) only occurs below a critical
temperature of 4 K. It is very difficult to imagine a scenario in which a
strong conventional superconducting order develops locally on such high
scales, but fully orders only at such low temperatures in a system that is
three dimensional, non-granular in structure, and not subjected to an
external magnetic field.

We will see that both these qualitative features of the problem are natural
consequences of the assumed existence of a PDW. Indeed, analogous features
have long been known to be a feature of the spin ordering in the same family
of materials\cite{kivelson03}. Specifically, unidirectional spin-stripe
order is observed to occur under a number of circumstances in the 214 family
of cuprate superconductors. However, when it occurs: 1) Although the
in-plane correlation length can be very long, in the range $100-400$\AA ,
the inter plane correlation length is never more than a few \AA , a degree
of anisotropy that cannot be reasonably explained simply\cite{sudip} on the
basis of the anisotropy in the magnitude of the exchange couplings. 2)
Despite the presence of long correlation lengths, true long-range
spin-stripe order has never been reported. It will be made clear, below,
that interlayer decoupling due to the geometry of the stripe order, and the
rounding of the ordering transitions by quenched disorder can be understood
as arising from closely analogous considerations applied to incommensurate
SDW and PDW order\cite{berg07}.

\section{Microscopic considerations}

\label{microscopic}

\begin{figure}[t]
\centering
\includegraphics[width=6cm]{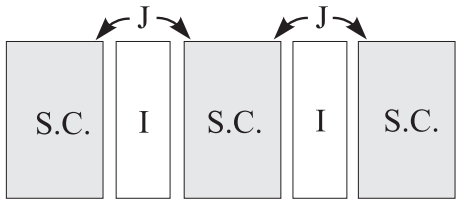}
\caption{Schematic model of a striped superconductor consisting of an array
of superconducting regions (S.C.) separated by correlated insulator regions
(I). The Josephson coupling between neighboring superconducting regions is $%
J $. If $J<0$, the system forms a striped superconducting (or PDW) state.}
\label{stripedSC}
\end{figure}

At first, the notion that a {PDW\ } phase could be stable sounds absurd.
Intuitively, the superconducting state can be thought of as the condensed
state of charge 2e bosons. However, in the absence of magnetic fields, the
ground-state of a bosonic fluid is always node-less, independent of the
strength of the interactions, and therefore cannot support a state in which
the superconducting order parameter changes sign. Thus, for a {PDW\ } state
to arise, microscopic physics at scales less than or of order the pair-size,
$\xi_0$, must be essential. This physics reflects an essential difference
between superfluids of paired fermions and preformed bosons\cite%
{SpivakKivelson}.

Our goal in this section is to shed some light on the mechanism by which
strongly interacting electrons can form a superconducting ground-state with
alternating signs of the order parameter. We will consider the case of a
unidirectional (striped) superconductor, but the same considerations apply
to more general forms of PDW order. We will not discuss the origin of the
pairing which leads to superconductivity. Likewise, we will not focus on the
mechanism of translation symmetry breaking by the density wave, as that is
similar to the physics of CDW and SDW formation. Our focus is on the \emph{%
sign alternation} of $\phi$. Thus, in much of this discussion, we will adopt
the model shown schematically in Fig. \ref{stripedSC}, in which we have
alternating stripes of superconductor and correlated insulator. The system
looks like an array of extended superconductor-insulator-superconductor
(SIS) junctions, and we will primarily be concerned with computing the
Josephson coupling across the insulating barriers. If the effective
Josephson coupling is positive, then a uniform phase (normal)
superconducting state is favored, but if the coupling is negative (favoring
a $\pi$ junction), then a striped superconducting phase is found.

\subsection{Previous results}

So long as time reversal symmetry is neither spontaneously nor explicitly
broken, the Josephson coupling, $J$ between two superconductors must be
real. If it is positive, as is the usual case (for reasons that will be
expanded upon in later subsections), the energy is minimized by the state in
which the phase difference across the junction is 0; if it is negative, a
phase difference of $\pi$ is preferred, leading to a ``$\pi$ junction.'' $%
\pi $ junctions have been shown, both theoretically and experimentally, to
occur for two distinct reasons: they can be a consequence of strong
correlation effects\cite%
{SpivakKivelson,otherpitheories,pijunctionsexperiments} or magnetic ordering%
\cite{pifrommagnetism1,pifrommagnetism2} in the junction region between two
superconductors, or due to the internal structure ({\ \textit{e.g.\/} }
d-wave symmetry) of the superconductors, themselves\cite%
{vanharlingen,kirtley}.

In the present paper, we will build on the first set of ideas. Until now, $%
\pi$ junctions have been confined to systems in which the Josephson
tunneling is dominated by a single impurity site or quantum dot. It is
non-trivial to extend this mechanism to the situation in which there is an
extended barrier in which $J$ is proportional to the cross-sectional
``area'' of the junction\cite{beasley}. Indeed, as we shall see below, under
most circumstances, even in the case in which the Josephson tunneling
through an isolated impurity would produce a negative $J$, for a barrier
consisting of an extended area of identical impurities, $J$ will typically
be positive. One thing that we achieve below is to obtain a proof that there
are circumstances in which $J$ of such an extended barrier is negative, and
to elucidate the conditions under which this occurs.

In the context of the cuprates, there have been several studies looking for
a striped superconducting state in the $t-J$ or Hubbard models. On the one
hand, DMRG calculations by White and Scalapino\cite{whiteandscalapino} have
consistently failed to find evidence in support of any sort of spontaneously
occurring $\pi$ junctions. On the other hand, a number of variational Monte
Carlo calculations have concluded that the striped superconductor is either
the ground-state of such a model\cite{ogata02}, under appropriate
circumstances, or at least close in energy to the true ground state\cite%
{poilblanc08a,poilblanc08b}. These variational calculations are certainly
encouraging, in the sense that they suggest that there is no large energetic
reason to rule out the existence of spontaneously occurring PDW order in
strongly correlated electronic systems. However, since no such state has yet
been observed in DMRG or other ``unbiased'' studies of these models, we
believe the mechanism of formation of these states, and indeed whether they
occur at all in physically reasonable microscopic models, remain unsolved
problems. It is these problems that we aim to address.

\subsection{$\protect\pi$ junctions from d-wave symmetry}

\label{dwave}

\begin{figure}[t]
\centering
\includegraphics[width=9cm]{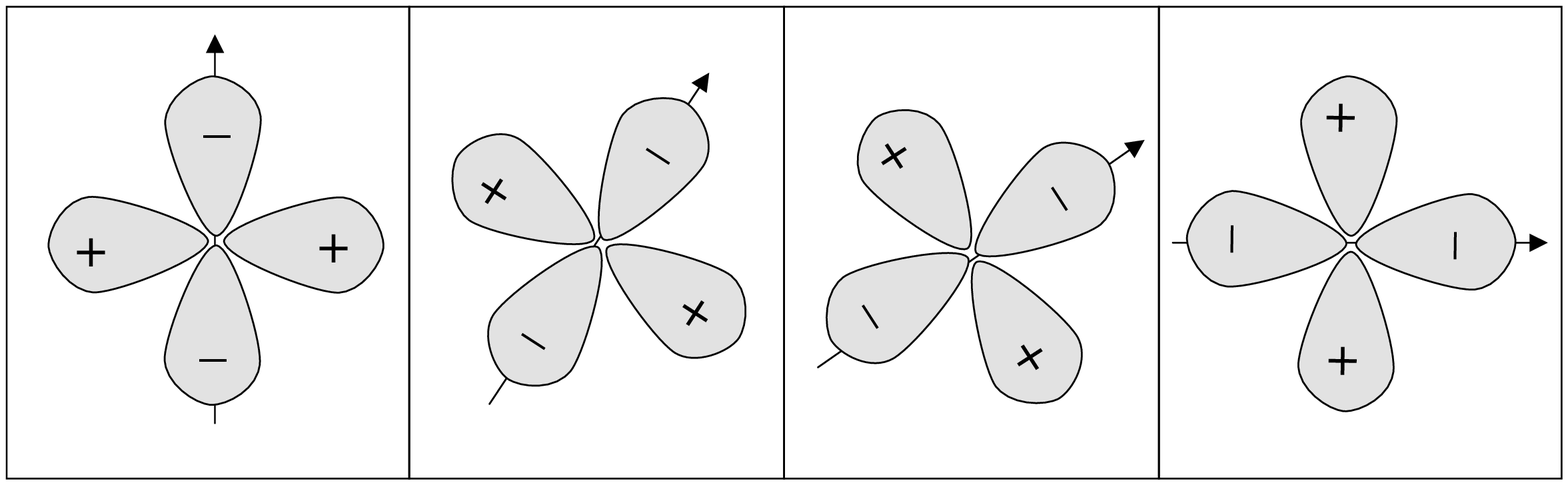}
\caption{An example of a $\protect\pi$ junction due to a rotating domain
walls in a system with a d-wave order parameter. From left to right, the
crystallographic axes rotates by $\protect\pi/6$ across each domain wall.
After three domain walls, the crystal axis has returned to itself, but the
order parameter has changed sign.}
\label{dwavestripes}
\end{figure}

In the case in which the preferred superconducting state has a non-trivial
internal pair-structure, as in a d-wave superconductor, it is well known
both theoretically and experimentally that $\pi $ junctions can be achieved
by suitable orientation of the crystal fields across grain boundaries. This
is \textit{not} the physics we have in mind. However, since it is simple and
well understood, we start with a highly artificial model in which d-wave
symmetry gives rise to a striped superconductor.

We consider the case of a d-wave superconductor in a square lattice, with a
strong crystal field coupling which locks the lobes of the pair
wave-function along the crystallographically defined x and y directions, as
shown in Fig. \ref{dwavestripes}. We imagine we have made a striped array in
which the crystallographic axis rotates by $\pi /6$ between neighboring
strips. Since rotation by $\pi /2$ is a symmetry, every third strip is
identical. However, the d-wave order parameter changes sign under this
rotation. Consequently, the ground-state is a striped superconductor in
which the period of the superconducting order parameter is twice the period
of the lattice structure.

This is a very contrived example, in which the origin of the {PDW} does not
require any strong coupling effects - it is derived from the internal
structure of the pair-wave function. For the point of principle, \textit{i.e.%
} to establish the viability of a {PDW} as a phase of matter, this analysis
is useful. However, for practical applications, a less contrived mechanism
is needed.

\section{Solved models}

\label{models}

\subsection{Model of a single junction}

To start with, we consider a system consisting of three strips - a ``left''
superconductor described by a mean-field Hamiltonian with gap function $%
\Delta _{L}(\vec{k})$ (not to be confused with the order parameter of Eq.\ %
\eqref{eq:op-delta})and quasiparticle energy $E_{L}(\vec{k})=\sqrt{[\epsilon
_{L}(\vec{k})]^{2}+|\Delta _{L}(\vec{k})|^{2}}$, a ``right'' superconductor
which is also described by a mean-field Hamiltonian with the index $L$
replaced by $R$, and between them a strongly correlated insulating ``
barrier'' region. Here $\vec{k}$ is either a 1D vector (if this is a line
junction) or a 2D vector if this is a planar junction. Note that we have
assumed the left and right superconductors are thin in the transverse
direction - otherwise we would have to include a transverse band index, as
well. We assume that both superconductors, by themselves, preserve
time-reversal symmetry, so in both cases the phase of the superconducting
order (modulo $\pi $) can be defined so that $\Delta_{\alpha }(\vec{k}%
)e^{-i\theta _{\alpha }}$ is real.

The three decoupled subsystems are thus described by the Hamiltonian
\begin{equation}
H_{0}=H_{L}+H_{B}+H_{R},
\end{equation}%
where, in the discussion below, we will consider several different examples
for the insulating barrier. However, in all cases, we will assume that the
superconducting Hamiltonians have the quadratic forms
\begin{eqnarray}
H_{\alpha } &=&\sum_{\vec{k}\sigma }\epsilon _{\alpha }(\vec{k})a_{\alpha ,%
\vec{k},\sigma }^{\dagger }a_{\alpha ,\vec{k},\sigma }  \notag \\
&&+\sum_{\vec{k}\sigma } \left[ \Delta _{\alpha }(\vec{k})a_{\alpha,\vec{k}%
,\sigma }^{\dagger }a_{\alpha ,-\vec{k}\sigma }^{\dagger }+\mathrm{h.c.}%
\right]
\end{eqnarray}%
with $\alpha =L$ or $R$ in which
\begin{equation}
a_{\alpha ,\vec{k},\sigma }=N^{-1/2}\sum_{\vec{r}}e^{i\vec{k}\cdot \vec{r}%
}c_{\alpha ,\vec{r},\sigma }
\end{equation}%
where $c_{\alpha ,\vec{r},\sigma }^{\dagger }$ creates an electron of spin
polarization $\sigma $ at position $\vec{r}$ on, respectively, the left
superconductor, the barrier, the right superconductor, for $\alpha =L,\ B$,
and $R$, and $N=\sum_{\vec{r}}1$ is the ``area'' of the junction.

The three subsystems are coupled together by a single-particle hopping term,
\begin{equation}
H^{\prime }=-\sum_{\vec{r},\sigma }[t_{L}c_{L,\vec{r},\sigma }^{\dagger }c_{%
\vec{r},\sigma } +t_{R}c_{R,\vec{r},\sigma }^{\dagger }c_{\vec{r},\sigma }+%
\mathrm{h.c.}]
\end{equation}%
(with the convention that $c_{\vec{r},\sigma }\equiv c_{B,\vec{r},\sigma }$
are the electron annihilation operators in the barrier region). Our purpose
is to determine how the ground-state energy depends on the difference in
phase between the superconducting order parameter on the left and right side
of the barrier, $\theta _{R}-\theta _{L}$. Specifically, we will consider
the limit in which the matrix elements in $H^{\prime }$ tend to zero, so
that the leading $\theta $ dependent term in the ground-state energy can be
computed in fourth order perturbation theory,
\begin{equation}
{\Delta E}=-2N{(t_{L}t_{R})^{2}}\left[ J\cos (\theta _{L}-\theta
_{R})+\ldots \right]
\end{equation}%
where $\ldots \rightarrow 0$ as $|t_{\alpha }|\rightarrow 0$, and $%
2|t_{R}t_{L}|^2J$ is the Josephson coupling density across the barrier.

An explicit expression can be obtained for $J$ in terms of imaginary
time-ordered correlation functions, as was shown in Ref. [%
\onlinecite{vadimandted}]. This can be derived either by making a Laplace
transform of the ordinary perturbative expression, or directly from an
Euclidean path integral:
\begin{equation}
J=\frac{1}{N\beta }\int \ d1\ d2\ d3\ d4\ F_{L}(1,2)F_{R}^{\star }(4,3)%
\tilde{\Gamma}(1,2;3,4)  \label{J}
\end{equation}%
where $1\equiv (\tau _{1},\vec{r}_{1})$,
\begin{equation}
\int d1\equiv \sum_{\vec{r}_{1}}\int_{0}^{\beta }d\tau _{1}
\end{equation}%
(in the limit $\beta \rightarrow \infty $) and
\begin{eqnarray}
F_{\alpha }(1,2) &\equiv & \left\langle T_{\tau }\left[ c_{\alpha ,\vec{r}%
_{1},\uparrow }^{\dagger }(\tau _{1}) c_{\alpha ,\vec{r}_{2},\downarrow}^{%
\dagger }(\tau _{2})\right] \right\rangle \\
\tilde{\Gamma}(1,2;4,3) &\equiv & \left\langle T_{\tau }\left[ c_{\vec{r}%
_{1},\uparrow }^{\dagger }(\tau _{1})c_{\vec{r}_{2},\downarrow
}^{\dagger}(\tau_{2}) c_{\vec{r}_{3},\downarrow }(\tau _{3})c_{\vec{r}%
_{4},\uparrow}(\tau _{4})\right] \right\rangle  \notag
\end{eqnarray}%
It is convenient to express $\tilde{\Gamma}$ as the sum of a non-interacting
piece, corresponding to processes in which two electrons tunnel through the
barrier independently of each other, plus a correction term, $\Gamma $,
which expresses the effects of interactions \textit{within the barrier region%
} between the tunneling electrons:
\begin{equation}
\tilde{\Gamma}(1,2;3,4)=G_{\uparrow }(1,4)G_{\downarrow
}(2,3)+\Gamma(1,2;3,4)  \label{tildegamma}
\end{equation}%
where
\begin{equation}
G_{\sigma }(1,2)\equiv \left\langle T_{\tau }\left[ c_{\vec{r}%
_{1},\sigma}^{\dagger }(\tau _{1})c_{\vec{r}_{2},\sigma }(\tau _{2})\right]
\right\rangle \text{.}
\end{equation}

Finally, we define two contributions to $J$:
\begin{equation}
J=J_1+J_2  \label{J1J2}
\end{equation}
where $J_1$ is the non-interacting portion, and $J_2$ is the portion
proportional to $\Gamma$.

\subsection{General considerations}

Under most circumstances, $J_{1}>0$ since it is proportional to $F^{2}G^{2}$
and hence does not depend on the sign of $G$ (see [\onlinecite{foot6}]). In
Ref. \onlinecite{vadimandted}, the situation in which there are attractive
interactions between electrons in the barrier was investigated. In this
case, $J_{2}$ is also, generally, positive. This earlier study dealt with
the conditions under which $J_{2}>J_{1}$, leading to anomalously large
values of the $I_{c}R$ product of the junction. In a sense, we are
investigating the converse problem, in which repulsive interactions lead to $%
J_{2}<0$, while at the same time $|J_{2}|>J_{1}$.

In most cases with conventional superconductors, in which $|\Delta |$ is the
smallest energy scale and $\xi _{0}$ (the superconducting coherence length)
is the largest length scale in the problem, it is typically the case that $%
J_{1}\gg |J_{2}|$. The dominant processes that contribute to $J_{1}$ involve
pairs tunneling within an (imaginary) time interval of each other $%
\sim1/|\Delta |$ and within a radius $\sim \xi _{0}$ of each other. In
contrast, the processes that dominate $J_{2}$ involve electrons tunneling
within an `` interaction time'' $\tau _{int}$ and an interaction range, $%
r_{int}$ of each other. Thus, one generally expects $|J_{2}/J_{1}|$ is small
in proportion to a positive power of $|\Delta |\tau_{int}$ and $r_{int}/\xi
_{0}$. In conventional superconductors, where these factors are very small,
this is the deciding factor, so that except under very special
circumstances, one expects positive (``ferromagnetic'') coupling across any
junction. Indeed, as discussed in Ref. \onlinecite{vadimandted}, in the
limit that $\Delta$ is small, Eq. \eqref{J} reduces to the familiar form for
a tunnelling Hamiltonian with an effective hopping $t_{\mathrm{eff}%
}=t_{L}gt_{R}$ across the barrier, where
\begin{equation}
g\equiv \int \frac{d1\ d2}{N\beta }G_{\sigma }(1,2).
\end{equation}%
Conversely, in ``high temperature superconductors,'' in which $\Delta $ is
not all that small, it is reasonable to expect, at least under some
circumstances, that the coupling is negative -- the long sought $\pi $%
-junction.

There is one other factor that can suppress $J_1$ relative to $J_2$: When
the two electrons tunnel across the barrier independently of each other,
their individual crystal momenta are conserved in the process. Thus, if we
remove a pair of electrons at momentum $\vec k$ and $-\vec k$ from the right
superconductor, they must be injected into the left superconductor at the
same momenta. If the right and left superconductors are the same, this is
not a problem -- electrons can be removed from near the Fermi energy on the
left and added on the right at low energy. However, if the left and right
superconductors have different values of $k_F$, then no such low energy
process is possible.
Conversely, for correlated tunneling, only the center of mass momentum of
the tunnelling electrons is conserved, so a pair of electrons removed from
the left with momenta $\vec k$ and $-\vec k$, can be injected into the right
superconductor with different momenta $\vec q$ and $-\vec q$.

\subsection{Explicit model ``barriers''}

\label{explicit-barriers}

To make the considerations explicit, we will compute $J$ for a model of the
barrier in several limits. Specifically, we will consider the case in which
the barrier consists of a Hubbard chain:
\begin{eqnarray}
H_{B} &=&\sum_{\vec{r}}\epsilon \hat{n}(\vec{r})-t\sum_{<\vec{r},\vec{r}%
^{\prime }>,\sigma } \left[ c_{\vec{r},\sigma }^{\dagger }c_{\vec{r},\sigma
}+\mathrm{h.c.}\right]  \notag \\
&&+U\sum_{\vec{r}}c_{\vec{r},\uparrow }^{\dagger }c_{\vec{r}%
,\downarrow}^{\dagger} c_{\vec{r},\downarrow }c_{\vec{r},\uparrow }
\end{eqnarray}%
in the strong coupling limit, where the hopping matrix element $t\rightarrow
0$, with site energy $\epsilon $ and on-site repulsion between two electrons
on the same site, $U>0$. We will consider two cases -- the empty chain case
in which $\epsilon >0$, and the half-filled case, in which $\epsilon
<0<U-\left\vert \epsilon \right\vert $. In the half-filled case, the
groundstate of the barrier is $2^{N}$ fold degenerate; we resolve this
degeneracy by applying a staggered magnetic field, $h\sum_{\vec{r},\sigma
}e^{i \vec{Q} \cdot \vec{r}}\sigma c_{\vec{r},\sigma }^{\dagger }c^{%
\vphantom{\dagger}}_{\vec{r},\sigma }$ where $\vec{Q}$ is the N\'{e}el
ordering vector, and we take the limit $h\rightarrow 0$ at the end of the
calculation.

For simplicity, we will also assume that $\Delta_L$ and $\Delta_R$ are
independent of $\vec k$, and that over the relevant range of energies, the
density of states in both superconductors can be approximated by a constant,
$\rho_L$ and $\rho_R$, respectively.

\subsubsection{The empty barrier}

\label{empty-barrier}

For the case of an empty barrier, $\epsilon > 0$, a straightforward (but
tedious) calculation reveals that
\begin{eqnarray}
J_1=&& \frac {2} N \sum_{\vec k} \left (\frac {\Delta_{L,\vec k}} {2
E_{L,\vec k}}\right ) \left (\frac {\Delta_{R,\vec k}} {2 E_{R,\vec k} }%
\right ) \left [ \frac 1 {E_{L,\vec k} + \epsilon}\right ] \left [\frac 1
{E_{R,\vec k} + \epsilon}\right ]  \notag \\
&&\times \left [ \frac 1 {E_{L,\vec k} + E_{R,\vec k}} + \frac 1 \epsilon
\right ].
\end{eqnarray}
If the left and right superconductors are the same
\begin{eqnarray}
J_1 =&& \frac \rho {2\epsilon\Delta} \left [ \frac{f(y)-y } {1-y^2}\right ].
\end{eqnarray}
where
\begin{eqnarray}
f(y_\alpha) = \frac {\cos^{-1}(y_\alpha)} {\sqrt{1-y_\alpha^2}},  \notag
\end{eqnarray}
$y_\alpha \equiv \epsilon/|\Delta_\alpha|$, and in this case $y=y_L=y_R$.
Here $f(y)$ is a smooth function with $f(0) = \pi/2$, $f(1) = 1$, and $f(y)
\sim \ln(2y)/y$ as $y\to\infty$.

If the two superconductors have quite different values of $k_F$, then $J_1$
gets substantially suppressed. Let $E_L$ and $E_R$ be, respectively, the
energy of the left quasiparticle at $k_{F,R}$ and the energy of the right
quasiparticle at $k_{F,L}$. So long as $E_L,\ E_R \ll \Delta$, the magnitude
of $J_1$ is hardly changed. However, if the two superconductors are
sufficiently different that $E_L,\ E_R \gg \Delta$, then
\begin{equation}
J_1 = \rho_L f(y_L)\left (\frac {\Delta_R} {E_R^2\epsilon }\right ) +
\rho_Rf(y_R)\left (\frac {\Delta_L} {E_L^2\epsilon }\right ).
\end{equation}

The interaction correction derives from terms that would have contributed to
$J_1$, but are suppressed due to the repulsive interaction, and so has the
opposite sign:
\begin{equation}
J_2 = -2 f(y_L) f(y_R)\rho_L\rho_R \left [ \frac U {\epsilon(U+2\epsilon)}
\right].
\end{equation}
In conventional superconductors, typically the band-width, $W$, is the
largest energy in the problem, and hence $\rho \sim W^{-1}$ is ``small.''
Since $J_1 \sim \rho$ and $J_2 \sim \rho^2$, this means that $J_1 \gg |J_2|$%
. The exception to this rule occurs in cases in which $\Delta E_F \equiv%
\mathrm{min}[E_R,E_L]$ is a substantial fraction of the bandwidth.

However, in high temperature superconductors, we expect to find that $\Delta
$ is not too much smaller than $W$. In this case, it will depend on details
whether or not $J_{1}$ or $J_{2}$ dominate. Clearly, this situation is more
likely the larger $U$. In the case of tunnelling between two identical
superconductors, and in the limit $U\rightarrow \infty$, $|J_{2}|>J_{1} $ so
long as
\begin{equation}
\rho \Delta >x_{c}\equiv \frac{\lbrack f(y)-y]}{4f^{2}(y)(1-y^{2})}.
\label{condition}
\end{equation}%
For $y\gg 1$, $x_{c}\sim y/\ln ^{2}y$, \textit{i.e.} condition %
\eqref{condition} requires impossibly large gap scales. However, $%
x_{c}=19/96 $ for $y=1$ and $x_{c}\rightarrow 1/2\pi $ as $y\rightarrow 0$.
\cite{foot7} Manifestly, $J_{1}$ is reduced in magnitude and $J_{2}$ is
unaffected by a substantial value of $\Delta E_{F}$. For instance, if we
consider the case in which the left and right superconductors differ only in
the position of the Fermi surface, $J_{2}$ dominates so long as
\begin{equation}
\lbrack \rho \Delta E_{F}]^{2}f(y)>\rho \Delta
\end{equation}

\subsubsection{The antiferromagnetic barrier}

\label{antiferromagnetic}

\begin{figure}[t]
\centering
\includegraphics[width=7cm]{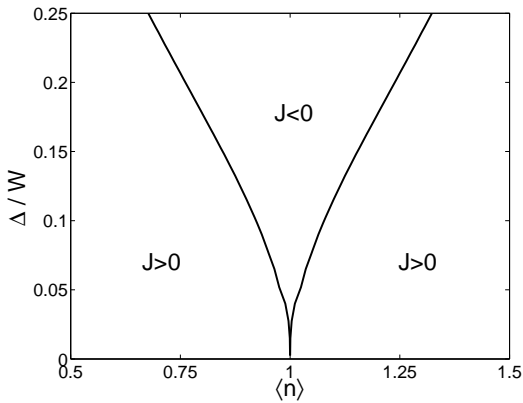}
\caption{Phase diagram for the antiferromagnetic barrier problem, as a
function of the pairing gap $\Delta$ (in units of the bandwidth $W=4t$) and
the average density $\langle n\rangle$ in the superconductors on either side
of the barrier. When $\Delta$ is large or $\langle n\rangle$ is close to 1
(half filling), then a negative Josephson coupling ($J<0$) is favored.}
\label{pdiagram}
\end{figure}

We now turn to discuss the case $\epsilon <0<U-\left\vert \epsilon
\right\vert $, in which the barrier is half filled. As mentioned previously,
we introduce a staggered Zeeman field $h$ to lift the groundstate
degeneracy, and we take $h\rightarrow 0$ at the end of the calculation. The
lowest order perturbative term in the Josephson coupling consists of all
processes in which a pair of electrons tunnels through the barrier, subject
to the constraint that the barrier has to return to its Ne\'{e}l ordered
ground state at the end of the process.

For simplicity, we consider the $U\rightarrow \infty $ case, in which no
doubly occupied sites are allowed in the barrier. (We have obtained similar
expressions for finite $U$, but they contain no qualitatively different
physics.) In this case, processes in which electrons tunnel though different
sites contribute to $J_{1}$ and those in which they tunnel through the same
site contribute to $J_{2}$. As discussed in Ref. \onlinecite{SpivakKivelson}%
, the single site process necessarily involves an exchange of two fermions,
and so makes a negative contribution to $J$. A new feature of the present
problem is that there an antiferromagnetic ``Umklapp''\ contribution to $%
J_{1}$, in which the tunneling electron exchanges momentum $\pi /a$ with the
antiferromagnet; this makes a negative contribution to $J_{1}$ which, under
appropriate circumstances, can largely cancel the usual positive
contribution.

It is a straightforward exercise to obtain the resulting expressions:
\begin{eqnarray}
&&J_{1}=\frac{1}{2N}\sum_{\vec{k},\vec{k}^{\prime }} \left( \frac{\Delta _{L,%
\vec{k}}}{2E_{L,\vec{k}}}\right) \left( \frac{\Delta _{R,\vec{k}^{\prime }}}{%
2E_{R,\vec{k}^{\prime }}}\right) \left( \delta _{\vec{k}-\vec{k}^{\prime}}
-\delta _{\vec{k}-\vec{k}^{\prime }+\pi \hat{y}}\right)  \notag \\
&&\times \left[ \frac{1}{E_{L,\vec{k}}+E_{R,\vec{k}^{\prime }}}+ \frac{1}{%
|\epsilon |}\right] \left[ \frac{1}{E_{L,\vec{k}}+|\epsilon |}\right] \left[
\frac{1}{E_{R,\vec{k}^{\prime }}+|\epsilon |}\right]  \notag \\
&&  \label{Ja}
\end{eqnarray}%
where $\delta _{\vec{k}}$ is the Kronecker delta function, and
\begin{eqnarray}
J_{2} &=&-\frac{1}{N^{2}}\sum_{\vec{k},\vec{k}^{\prime }}\left( \frac{\Delta
_{L,\vec{k}}}{2E_{L,\vec{k}}}\right) \left( \frac{\Delta _{R,\vec{k}%
^{\prime}}}{2E_{R,\vec{k}^{\prime }}}\right)  \label{Jb} \\
&&\times \left[ \frac{1}{E_{L,\vec{k}}+|\epsilon |}\right] \left[ \frac{1}{%
E_{R,\vec{k}^{\prime }}+|\epsilon |}\right] \left[ \frac{1}{E_{L,\vec{k}}
+E_{R,\vec{k}^{\prime }}}\right] .  \notag
\end{eqnarray}%
Generically, the magnitude of the umklapp term (proportional to $\delta _{%
\vec{k}-\vec{k}^{\prime }+\pi \hat{y}}$) is considerably smaller than the
momentum conserving term, since it is typically not possible for both $\vec{k%
}$ or $\vec{k}+\pi $\emph{\^{y} } to be close to the Fermi surface \emph{%
unless} a special ``nesting'' condition is satisfied. For example, if the
two superconductors are one dimensional wires close to half filling, then
the separation between the two Fermi points is indeed close to $\pi $. In
that case, the umklapp term tends to cancels the momentum conserving term.
(If particle-hole symmetry is present, this cancellation is exact.)

We have evaluated the expressions in Eqs. \eqref{Ja} and \eqref{Jb}
numerically, for the case in which the two superconducting strips are
identical chains with a $k$ independent gap function, $\Delta
_{\alpha,k}=\Delta $ and the band-structure is the simplest tight-binding
result, $\epsilon _{\alpha ,k}=-2t\cos (k)-\mu $. The results are summarized
in Fig. \ref{pdiagram}, where the x axis is the average density of electrons
$\langle n\rangle $ in the superconductors, and the y-axis is the magnitude
of $\Delta $; the solid lines separate the region of net negative from the
regions of net positive Josephson coupling. The general trends alluded to in
the above discussion are clear from this figure.

\subsection{A model with a striped superconducting ground state}

\label{model-striped_sc}

The basic ingredients necessary for finding a model with a striped
superconducting ground-state are present in the above discussed examples.
However, in the interest of making everything explicit, we consider the
following problem. Firstly, we generalize the problem to $d+1$ dimensions.
We consider an array of parallel $d$ dimensional hyper-planes with weak
attractive interactions, and a particle-hole symmetric dispersion. Each
hyperplane thus has a uniform superconducting order parameter. For weak
attractions between electrons and large enough $d$, the superconducting
correlations in each hyperplane can, presumably, be accurately treated in
the context of BCS mean-field theory. The particle-hole symmetry implies
that $E_{\vec k} = E_{\vec k +\vec \pi}$, where $\vec \pi$ is the $d$%
-dimensional antiferromagnetic N\'eel ordering vector. Sandwiched between
each superconducting hyperplane there is a $d$-dimensional insulting
hyperplane, which is the $d$-dimensional generalization of the
antiferromagnetic barrier discussed above. Finally, we couple neighboring
hyperplanes with an arbitrarily weak hopping matrix, $t$. The Josephson
coupling per unit hyperarea is then computed as above.

The particle-hole symmetry of the model insures that $J_{1}=0$, and that
therefore the Josephson coupling per unit hyper-area is proportional to $%
J_{2}$, and hence is negative. Thus, the ground-state of this system is a
state with coexisting SDW and striped superconducting order. Note that the
phase of the antiferromagnetic order is determined by the sign of $h$. If we
take $h>0$ in all cases, then the antiferromagnetic order is ``in-phase.''
In this case, the Hamiltonian is invariant under translations in the
direction perpendicular to the hyperplanes by the distance between
insulating hyperplanes, and hence the striped superconductor spontaneously
breaks translational symmetry. However, we can also imagine staggering the
sign of $h$ from one hyperplane to the next, so as to mimic the antiphase
spin-stripe order seen in the cuprates. In this case, the underlying unit
cell is twice as large, and the striped superconductor does not
spontaneously break translational symmetry. However, in both cases, the
spatial average of any component of the superconducting order parameter is
0, so this state conforms to our definition of a PDW.

This model is, admittedly, somewhat contrived, especially in the assumed
particle-hole symmetry. However, since at this point $J_{2}$ has a finite
magnitude, the state is robust with respect to small deformations of the
model.

The construction of solvable microscopic models with a striped
superconducting ground-state, and the elucidation of the ingredients
necessary to get anti-phase superconducting order across a correlated
insulating barrier, are the principle results of the present paper.

\subsection{Quasiparticle spectrum of a striped superconductor}

\label{qp-spectrum}

The spectrum of an s-wave superconductor is typically fully gapped, while a
d-wave superconductor has a gap that vanishes only on a few isolated points
(lines in 3d) on the Fermi surface. These features persist even in the
presence of coexisting CDW or SDW order of various sorts\cite{bergandchen}.
In particular, the density of quasiparticle states is either zero for
energies less than a non-zero minimum value (in the s-wave case), or
vanishes (linearly) as the energy approaches the Fermi surface. This effect
of superconducting order is in contrast to most other orders, including SDW
and CDW order, in which, so long as the effective ordering potential is weak
compared to the Fermi energy, at most a portion of the Fermi surface is
gapped leaving the system with a (smaller) reconstructed Fermi surface, but
a Fermi surface none-the-less.

The gap character in a PDW is more similar to that of a CDW than to a
uniform superconductor. Specifically, so long as the order parameter is not
too large, an ungapped reconstructed Fermi surface remains\cite%
{bergandchen,dror}. Thus, a PDW typically has a finite density of states in
the superconducting phase.

For band-structure parameters characteristic of a typical underdoped or
optimally doped curpate, the quasiparticle spectrum of a striped
superconductor has a large (d-wave-like) gap near the `` antinodal points''\
(\textit{i.e.} near $(\pi ,0)$ and $(0,\pi )$), but has a closed,
Fermi-surface pocket which closely follows the contours of the portion of
the normal-state Fermi surface near the `` nodal points''\ [\textit{i.e.}
where the Fermi surface crosses the $(0,0)$ to $(\pi ,\pi)$ cord]. In short,
it closely resembles certain phenomenological descriptions\cite{shenpg} of
the ``pseudo-gap.'' The low energy spectral weight is substantial only on a
portion of the bare Fermi surface near the nodal $(\pi ,\pi)$ direction\cite%
{sudip}, forming a ``Fermi arc'' \cite{Norman, Kanigel}.

\section{Coupled order parameters}

\label{op}

In this section, we explore the aspects of the theory of a {PDW} that can be
analyzed without reference to microscopic mechanisms. We focus on the
properties of ordered states at $T=0$, far from the point of any quantum
phase transition, where for the most part fluctuation effects can be
neglected. (The one exception to the general rule is that, where we discuss
effects of disorder, we will encounter various spin-glass related phases
where fluctuation effects, even at $T=0$, can qualitatively alter the
phases.) For simplicity, most of our discussion is couched in terms of a
Landau theory, in which the effective free energy is expanded in powers of
the order parameters; this is formally \textit{not} justified deep in an
ordered phase, but it is a convenient way to exhibit the consequences of the
order parameter symmetries.

Specifically, the emphasis in this section is on the interrelation between
striped superconducting order and other orders. There is a necessary
relation between this order and CDW and nematic (or orthorhombic) order,
since the striped superconductor breaks both translational and rotational
symmetries of the crystal. From the microscopic considerations, above, and
from the phenomenology of the cuprates, we also are interested in the
relation of superconducting and SDW order. We assume that the host crystal
is tetragonal, and that there are therefore two potential symmetry related
ordering wave-vectors, $Q$ and $\bar{Q}$, which are mutually orthogonal.
Spin-orbit coupling is assumed to be negligible. We introduce the following
order parameter fields: the nematic, $N$, the PDW $\Delta _{Q}$, the SDW $%
\vec{S}_{Q}$, and the CDW $\rho _{K}$ (where $K=2Q $), and, of course, the
corresponding orders at the symmetry related wave-vector, $\bar{Q}$. The
Landau effective free energy density can then be expanded in powers of these
fields:
\begin{equation}
\mathcal{F}=\mathcal{F}_{2}+\mathcal{F}_{3}+\mathcal{F}_{4}+\ldots
\end{equation}%
where $\mathcal{F}_{2}$, the quadratic term, is simply a sum of decoupled
terms for each order parameter,
\begin{eqnarray}
\mathcal{F}_{3} &=&\gamma _{s}[\rho _{K}^{\star }\vec{S}_{Q}\cdot \vec{S}%
_{Q} +\rho _{{\bar{K}}}^{\star }\vec{S}_{\bar{Q}}\cdot \vec{S}_{\bar{Q}}+%
\mathrm{c.c.}] \\
&&+\gamma _{\Delta }[\rho _{K}^{\star }\Delta _{-Q}^{\star }\Delta _{Q}
+\rho_{\bar{K}}^{\star }\Delta _{-\bar{Q}}^{\star }\Delta_{\bar{Q}}+\mathrm{%
c.c.}]  \notag \\
&&+g_{\Delta }N[\Delta _{Q}^{\star }\Delta _{Q}+\Delta _{-Q}^{\star }\Delta
_{-Q}-\Delta _{\bar{Q}}^{\star }\Delta _{\bar{Q}}-\Delta _{-\bar{Q}}^{\star
}\Delta _{-\bar{Q}}]  \notag \\
&&+g_{s}N[\vec{S}_{Q}^{\star }\cdot \vec{S}_{Q}-\vec{S}_{\bar{Q}}^{\star
}\cdot \vec{S}_{\bar{Q}}]  \notag \\
&&+g_{c}N[\rho _{K}^{\star }\rho _{K}-\rho _{\bar{K}}^{\star }\rho _{\bar{K}%
}],  \notag
\end{eqnarray}%
and the fourth order term, which is more or less standard, is shown
explicitly in Appendix \ref{app:appendixA}. (Detailed discussions of the
microscopic definition of the PDW order parameter is also discussed in this
Appendix.)

The effect of the cubic term proportional to $\gamma _{s}$ on the interplay
between the spin and charge components of stripe order has been analyzed in
depth in Ref. \onlinecite{zke}. Similar analysis can be applied to the other
terms. In particular, it follows that the existence of superconducting
stripe order ($\Delta _{Q}\neq 0$, and $\Delta _{\bar{Q}}=0$), implies the
existence of nematic order ($N\neq 0$) and charge stripe order with half the
period ($\rho _{2Q}\neq 0$). However, the converse statement is not true:
while CDW order with ordering wave-vector $2Q$ or nematic order tend to
promote PDW order, depending on the magnitude of the quadratic term in $%
\mathcal{F}_{2}$, PDW order may or may not occur.

One new feature of the coupling between the PDW and CDW order is that it
produces a sensitivity to disorder which is not normally a feature of the
superconducting state. In the presence of quenched disorder, there is always
some amount of spatial variation of the charge density, $\rho (r)$, of which
the important portion for our purposes can be thought of as being a pinned
CDW, that is, a CDW with a phase which is a pinned, slowly varying function
of position, $\rho (r)=|\rho _{K}|\cos [K\cdot r+\phi (r)]$. Below the
nominal striped superconducting ordering temperature, we can similarly
express the PDW order in terms of a slowly varying superconducting phase, $%
\Delta (r)=|\Delta _{Q}|\exp [iQ\cdot r+i\theta _{Q}(r)]+|\Delta_{-Q}|\exp
[-iQ\cdot r+i\theta _{-Q}(r)]$. The resulting contribution to $\mathcal{F}%
_{3}$ is
\begin{equation}
\mathcal{F}_{3,\gamma }=2\gamma _{\Delta }|\rho _{K}\Delta _{Q}\Delta
_{-Q}|\cos [2\theta _{-}(r)-\phi (r)].  \label{F3gamma}
\end{equation}%
where
\begin{eqnarray}
\theta _{\pm }(r) &\equiv &[\theta _{Q}(r)\pm \theta _{-Q}(r)]/2;
\label{thetapm} \\
\theta _{\pm Q}(r) &=&[\theta _{+}(r)\pm \theta _{-}(r)]\text{.}  \notag
\end{eqnarray}%
The aspect of this equation that is notable is that the disorder couples
directly to a piece of the superconducting phase, $\theta _{-}$. No such
coupling occurs in usual 0 momentum superconductors.

It is important to note that the condition that $\Delta (r)$ be single
valued imposes a non-trivial topological constraint on possible vortices.
Specifically, an isolated half-vortex (about which the phase winds by $\pi $%
) is forbidden in either $\theta _{+}$ or $\theta _{-}$; vortices must occur
either as full ($2\pi $)-vortices in one or the other phase field, or as a
bound pair of a $\theta _{+}$ and a $\theta _{-}$ half-vortex. An important
consequence of this phase coupling is that the effect of quenched disorder,
as in the case of the CDW itself, destroys long-range superconducting stripe
order. (This statement is true\cite{larkin}, even for weak disorder, in
dimensions $d<4$.) Naturally, the way in which this plays out depends on the
way in which the CDW state is disordered.

In the most straightforward case, the CDW order is punctuated by random,
pinned dislocations, \textit{i.e.} $2\pi $ vortices of the $\phi$ field. The
existence of the coupling in Eq. \eqref{F3gamma} implies that there must be
an accompanying $\pi $ vortex in $\theta _{-}$. The condition of
single-valued-ness implies that there must also be an associated half-vortex
or anti-vortex in $\theta _{+}$\cite{tsunetsugu}. If these latter vortices
are fluctuating, they destroy the superconducting state entirely, leading to
a resistive state with short-ranged striped superconducting correlations. If
they are frozen, the resulting state is analogous to the ordered phase of an
XY spin-glass: such a state has a non-vanishing Edwards-Anderson order
parameter, spontaneously breaks time-reversal symmetry, and, presumably, has
vanishing resistance but no Meissner effect and a vanishing critical
current. In 2D, according to conventional wisdom, a spin-glass phase can
only occur at $T=0$, but in 3D there can be a finite temperature glass
transition\cite{youngreview}.

In 3D there is also the exotic possibility that, for weak enough quenched
disorder, the CDW forms a Bragg-glass phase, in which long-range order is
destroyed, but no free dislocations occur\cite{giamarchi,daniel,gingras}. In
this case, $\phi $ can be treated as a random, but single-valued function -
correspondingly, so is $\theta _{-}$. The result is a superconducting
Bragg-glass phase which preserves time reversal symmetry and, presumably,
acts more or less the same as a usual superconducting phase. It is believed
that a Bragg-glass phase is not possible in 2D\cite{daniel}.

A summary of the characterization of the PDW, SC and CDW phases in terms of
their order parameters, their inter-relations and their sensitivity to
quenched disorder appears in Table \ref{tab:phases}.

Another perspective on the nature of the superconducting state can be
obtained by considering a composite order parameter which is proportional to
$\Delta_Q\Delta_{-Q}$. There is a cubic term which couples a uniform, charge
4e superconducting order parameter, $\Delta_4$, to the PDW order:
\begin{equation}
\mathcal{F}_3^\prime =g_4 \{\Delta_4^\star [\Delta_Q \Delta_{-Q} +
\Delta_{\bar Q} \Delta_{-\bar Q}] +\mathrm{c.c.}\}
\end{equation}
This term implies that whenever there is PDW order, there is also
necessarily charge 4e uniform superconducting order. However, since this
term is independent of $\theta_-$, it would be totally unaffected by
Bragg-glass formation of the CDW. The half-vortices in $\theta_+$ discussed
above can simply be viewed as the fundamental ($hc/4e$) vortices of a charge
4e superconductor.

Turning now to the quartic terms in $\mathcal{F}_{4}$, there are a number of
features of the ordered phases which depend qualitatively on the sign of
various couplings. Again, this is very similar to what happens in the case
of CDW order - see, for example, Ref. [\onlinecite{robertson}]. For
instance, depending on the sign of a certain biquadratic term, either
unidirectional (superconducting stripe) or bidirectional (superconducting
checkerboard) order is favored.

Finally, we comment on the case of coexisting uniform and striped
superconducting order parameters. Such a state is not thermodynamically
distinct from a regular (uniform) superconductor coexisting with a charge
density wave, even if the uniform superconducting component is in fact
weaker than the striped component. Therefore, we expect many of the special
features of the striped superconductor (such as its sensitivity to potential
disorder) to be lost. In Appendix \ref{app:uniform}, we extend the Landau
free energy to include a uniform superconducting component, and show that
this is indeed the case.

\section{Model for a layered striped superconductor}

\label{glass_model} We now discuss a low energy effective theory for a
layered, disordered striped superconductor. As we saw in the previous
section, disorder inevitably nucleates half vortices which are pinned to
dislocations in the CDW. The system therefore has Ising-like degrees of
freedom which are the \textquotedblleft charges\textquotedblright\ (or
vorticities) of these $\pi $ vortices. We assume that their positions are
quenched, random variables. As we discussed in Ref. [\onlinecite{berg07}], a
PDW in a layered superconductor can give rise to a frustration of the
inter--layer Josephson coupling, given a structure in which the CDW\ is
rotated by $90^{\circ }$ between layers [as occurs in the low temperature
tetragonal (LTT) phase of La$_{2-x}$Ba$_{x}$CuO$_{4}$]. To make the model
relevant to this system, we consider the case where the inter--layer
Josephson coupling vanishes identically.

Under these assumptions, the interaction energy between half vortices is
composed of two parts: the magnetic energy and the kinetic energy associated
with the screening currents that surround the vortices. The total
interaction energy for a given configuration of half vortices is\cite%
{tinkham}
\begin{equation}
U=\frac{1}{2}\sum_{ij\alpha \beta }u\left( \vec{R}_{i\alpha }-\vec{R}%
_{j\beta }\right) q_{i\alpha }q_{j\beta }+\frac{1}{2}\sum_{\alpha \beta
}v\left( z_{\alpha }-z_{\beta }\right) Q_{\alpha }Q_{\beta }\text{.}
\label{Utot}
\end{equation}%
Here, $q_{i\alpha }=\pm 1$ and $\vec{R}_{i\alpha }$ are the
\textquotedblleft charge\textquotedblright\ (or vorticity) and the position
of the $i$th vortex in layer $\alpha $, respectively, $z_{\alpha }$ is the $%
z $ coordinate of layer $\alpha $, and $Q_{\alpha }=\sum_{i}q_{i\alpha }$.
The interaction potentials $u(\vec{R})$ and $v\left( z\right) $ are given by
\begin{equation}
u\left( \vec{R}\right) =\frac{B_{0}^{2}\lambda ^{2}d}{\mu _{0}}\int \frac{%
d^{2}k}{\left( 2\pi \right) ^{2}}\left[ \frac{\delta _{\alpha \beta }}{k^{2}}%
-\frac{d}{2\lambda }\frac{e^{-\frac{\alpha \left( k\right) \left\vert
z\right\vert }{\lambda }}}{\alpha \left( k\right) k^{2}}\right] \left[ e^{i%
\vec{k}\cdot \vec{\rho}}-1\right] \text{,}  \label{u}
\end{equation}%
where $\vec{\rho}$ and $z$ are the radial and the $z$ axis separation
between the two vortices, respectively, $\alpha \left( k\right) \equiv \sqrt{%
1+\left( \lambda k\right) ^{2}}$, $B_{0}\equiv \frac{h}{4e\lambda ^{2}}$, $%
\delta _{\alpha \beta }$ is a Kronecker delta function of the layer indices $%
\alpha ,\beta $ of the two vortices, $\lambda $ is the in-plane penetration
depth, $d$ is the inter-layer distance, and
\begin{eqnarray}
v\left( z\right) &=&\frac{B_{0}^{2}\lambda ^{2}d}{2\pi \mu _{0}}{\bigg \{}%
\delta _{\alpha \beta }\ln \frac{L}{\xi }-\frac{d}{2\lambda }e^{-\frac{%
\left\vert z\right\vert }{\lambda }}\ln \frac{L}{\lambda } \\
&&+\frac{d}{4\lambda }e^{-\frac{2\left\vert z\right\vert }{\lambda }}\left[
\ln \frac{\left\vert z\right\vert }{2\lambda }+\gamma +e^{\frac{2\left\vert
z\right\vert }{\lambda }}\mathrm{E}_{1}\left( \frac{2\left\vert z\right\vert
}{\lambda }\right) \right] {\bigg \}}\text{,}  \notag
\end{eqnarray}%
where $L$ is a long wavelength cutoff of the order of the linear dimension
of the system, $\xi $ is the coherence length, $\mathrm{E}_{1}\left(
x\right) =\int_{x}^{\infty }\frac{e^{-t}}{t}dt$ is the exponential integral
function, and $\gamma $ is Euler's constant. The second term in Eq. (\ref%
{Utot}) diverges as $\ln L$, unless $Q_{\alpha }=Q=\text{const}$. for all $%
\alpha $. In the absence of an external magnetic field, this term constrains
$Q_{\alpha }=0$. This is the usual \textquotedblleft charge
neutrality\textquotedblright\ condition which comes from the infrared
divergence of the vortex self energy.

For $\sqrt{\rho ^{2}+z^{2}}\gg \lambda $, Eq. (\ref{u}) reduces to
\begin{equation}
u\left( \vec{R}\right) \approx \frac{B_{0}^{2}\lambda ^{2}d}{2\pi \mu _{0}}%
\left[ \delta _{\alpha \beta }-\frac{d}{2\lambda }e^{-\frac{\left\vert
z\right\vert }{\lambda }}\right] \ln \frac{\xi }{\rho }\text{.}  \label{U}
\end{equation}%
The statistical mechanics problem of a finite density of half vortices with
quenched random positions, and whose interaction is given by Eq. (\ref{Utot}%
), is an interesting unsolved statistical mechanics problem.\cite%
{kumar,spinglass}

As we mentioned, the putative superconducting glass phase (in which the half
vortices are frozen) necessarily involves time reversal symmetry breaking.
It should therefore be detectable by measuring the magnetic fields
associated with the spontaneous half vortices which occur at dislocations in
the charge order. For example, consider a half vortex at the surface of the
sample (taken to be at $z=0$).
The asymptotic form of the magnetic field above the surface in the limit $%
0<z\ll \lambda $, is 
\begin{equation}
B_{z}\left( z\ll \lambda \right) \approx \frac{\hbar }{8\lambda ^{2}e}\left(
\frac{d}{z}\right) \text{,}  \label{Bz}
\end{equation}%
where $d$ is the interlayer distance. [For $z\gtrsim \lambda $, screening by
diamagnetic currents in the other planes becomes considerable, and $B_{z}$
crosses over to $\sim \frac{\hbar d}{8\lambda e}\left( \frac{1}{z^{2}}%
\right) $.] Assuming $\lambda \approx 2000$ \AA\ and $d\approx 15$ \AA , Eq. %
\eqref{Bz} gives $B_{z}\approx 300\left( \frac{\text{\AA }}{z}\right) $ G.
At a distance of about $1000$ \AA\ from surface, the resulting
characteristic fields of the order of $0.3$ G (assuming a half vortex right
at the surface) are well within the resolution of current local magnetic
field measurement techniques. The onset of a spontaneous, random magnetic
fields of this order at the glass transition temperature (which is
presumably also signaled by the vanishing of the linear resistivity) would
be a dramatic confirmation of the PDW scenario in $\frac{1}{8}$ doped LBCO.










\section{Final thoughts}

\label{final}

The strong suppression of the three-dimensional superconducting $T_c$ in {La$%
_{2-x}$Ba$_x$CuO$_4$} at $x=1/8$, and other ``$1/8$ anomalies'' in the {La$%
_{2-x}$Sr$_x$CuO$_4$} family of cuprate superconductors, have long been
interpreted in the literature as evidence that charge order competes with
high temperature superconductivity. (More recently\cite%
{bonnhardy,tailleferunp,ando}, clear evidence of a 1/8 anomaly has been
adduced in {YBa$_2$Cu$_3$O$_{7-\delta}$}, as well.) However, the remarkable
properties of {La$_{2-x}$Ba$_x$CuO$_4$}, particularly the fact that the
anti-nodal gap is largest at $x=1/8$, where the dynamical layer decoupling
is observed, strongly suggests that charge stripe order \emph{can be} part
of the mechanism of superconductivity as argued in Ref.[\onlinecite{afk}].

The superconducting state presented here represents a new face of the
interplay between superconductivity and spin and charge order. If the SC and
SDW orders simply competed, the PDW would be a rather unnatural state; it is
natural if self-organized inhomogeneities are an essential feature of the
mechanism. In this sense, the PDW state should also be regarded as an
electronic liquid crystal phase.\cite{kivelson98} Numerous and important
implications follow from the properties of this state.

\subsection{Implications}

\label{implications}


\emph{Glassy superconductors:} The most dramatic consequence of the PDW
physics is that, in the presence of weak, quenched disorder, the
superconducting phase gives way to a regime of glassy behavior, where strong
local superconducting correlations extend up to a finite correlation length
set by the strength of the disorder. In the regime of temperature below the
onset of substantial local superconducting coherence but above the
transition to a fully superconducting state (if one occurs), the system can
be characterized as a ``failed superconductor'' \cite{shen08}. Clearly, in
this regime, the longitudinal resistivity will be small compared to normal
state values, as will the quasiparticle contributions to the thermopower and
linear Hall resistance. Moreover, one expects to see strong local
indications of superconductivity, especially the formation of a
superconducting pseudo-gap in the single-particle spectrum with a form
similar to that of the ordered phase.

Taking into account that the materials are ultimately 3D, there will
generally be a glass transition at $T=T_{g}$ in this regime. In the absence
of an applied magnetic field, the glass phase can be characterized by its
spontaneously broken time-reversal symmetry. Presumably, the glass phase has
zero resistance, zero critical current, and fails to exhibit a full Meissner
effect. More generally, for a range of temperatures, including temperatures
above $T_{g}$, the magnetic response of the PDW will be characterized by
slow dynamics, a broad distribution of relaxation rates, and probably a
certain degree of history dependence; all the dramatic and confusing
features of spin-glasses, but amplified by the coherent orbital organization
of a superconducting state.

\emph{Interlayer decoupling:} When a PDW state occurs in a quasi-2D
(layered) material, it will frequently be the case (depending on the
interlayer geometry) that the usual Josephson coupling between neighboring
layers vanishes. This feature was explored at length in our earlier paper,
Ref. [\onlinecite{berg07}]. In the case of the LTT structure of {La$_{2-x}$Ba%
$_x$CuO$_4$}, the fact that the stripes in neighboring planes run at right
angles to each other insures that (in the absence of disorder), the
Josephson coupling between neighboring planes vanishes. Strictly speaking,
this does not mean that there is no coupling between plains, as would occur
in a putative ``floating phase'' \cite{lubensky}. There are always couplings
between farther neighbor layers, and higher order Josephson couplings
between neighboring planes (which couple $\Delta _{4}$ in 3D), and these
interactions are relevant in the renormalization group sense at any
temperature below the putative Kosterlitz-Thouless (KT) transition
temperature. However, at the very least, it means that the 3D
superconducting state is enormously more anisotropic than would be expected
on the basis of the bare electronic anisotropy. Moreover, as long as
superconducting coherence in a given plane is limited due to quenched
disorder, if these higher order couplings are weak enough, they can be
essentially ignored.

\emph{Striped superconductors in strongly correlated models:} While we have
established, as a point of principle, that well defined models exist that
support a PDW phase, it still makes sense to make predictions that can be
tested in ``numerical experiments'' on $t-J$ and Hubbard models. The fact
that spontaneously occurring $\pi $-junctions have not yet been reported in
extensive previous DMRG studies\cite{whiteandscalapino} is worrisome, in
this regard.

Based upon the insights gained in the above, we propose DMRG studies of
microscopic models in which we expect indications of PDW formation can be
observed. Two examples are:

i) A three leg Hubbard ladder in which the outer two legs have a negative $U$
and the inner-leg a positive $U$. The chemical potential and on-site
energies should be chosen so that the inner leg is near half-filling, and so
strongly antiferromagnetically correlated. The density of electrons on the
outer (``superconducting'') legs can be varied, and need not be identical.
This model is thus a close relative of the model that we treated in Sec.
VI.C.2, except that we do not treat the superconducting legs in mean-field
theory and we are not restricted to considering parametrically small
coupling between the legs. The tendency to a striped SC phase can be tested
by studying the sign of the pair-field correlations between the upper and
lower legs. When the outer legs are near half filling, we expect to see
negative correlations, indicative of PDW formation. Indeed, we have already
seen such behavior in preliminary DMRG studies of this model\cite{berg08b}.

ii) A five leg Hubbard ladder with all repulsive interactions, constructed
to consist of two outer two-leg ladders and an inner Hubbard chain. Again,
the chemical potential and on-site energies should be chosen so that the
inner chain is near half-filling, making it strongly antiferromagnetically
correlated, while the density of electrons on the outer ladders can be
varied, and need not be identical. By making the densities on the outer legs
different enough, we are confident that a PDW state can be induced. However,
we are very curious to learn how robust this state is under less contrived
conditions. We are now undertaking such an investigation\cite{berg08b}.

\subsection{Speculations}

\label{speculations}

\emph{Striped SC phases in {La$_{2-x}$Ba$_{x}$CuO$_{4}$} at $x=1/8$:} We
were originally motivated in this study by experiments in {La$_{2-x}$Ba$_x$%
CuO$_4$}, and we remain optimistic that they are, indeed, evidence of the
existence of a PDW phase. In order to make more than an impressionistic
comparison with experiment, the nature of the superconducting glass phase
will need to be understood theoretically, much better than we do at present.
However, in the absence of such theoretical control, we can still make some
speculative statements concerning the relation between experiment and theory.

Tentatively, we would like to identify the point at which resistivity
vanishes in the c-direction as the true point of the 3D glass transition, $%
T_g \approx 10$K. If this is the case, the linear-response resistivity at
lower temperatures is truly zero, although various non-linear and hysteretic
processes may complicate the measurements. We expect very long time-scales
and a degree of history dependence of macroscopic measurements to begin
being significant at considerably higher temperatures, $T > T_g$.

It is natural to associate the point at which the in-plane resistivity
apparently vanishes $T_{KT} \approx 16$K, with what would have been the KT
transition of a single plane in the absence of quenched disorder. We do not
think this is a true transition, and would expect that if experiments could
be carried out with higher precision than has currently been possible, the
in-plane resistivity would be found to have a finite value for $%
T_{g}<T<T_{KT}$. In relatively clean samples, simple scaling arguments
suggest that the residual resistivity should be proportional to $\xi^{-2}$,
where $\xi$ is the coherence length, which in turn is roughly the distance
between dislocations.

The sharp drop in the resistivity at the spin-ordering temperature, $%
T_{SDW}\approx42$K, is probably not a true phase transition, either, but
rather marks the sudden onset of significant intermediate scale
superconducting coherence. However, we suspect that local pairing
correlations persist to higher temperatures; as long as the spin order is
strongly fluctuating at $T > T_{SDW}$, we imagine that phase coherence
between neighboring superconducting stripes is prevented, while for $T <
T_{SDW}$, antiphase superconducting correlations extend over multiple
stripes. We favor this viewpoint for several reasons, most importantly
because the gap-features seen in ARPES persist to higher temperatures.\cite%
{shen08}

\emph{ARPES spectrum of {La$_{2-x}$Ba$_x$CuO$_4$}:} A recent study\cite%
{shen08} of the ARPES spectrum of {La$_{2-x}$Ba$_x$CuO$_4$} confirmed the
conclusion of an earlier\cite{valla06} ARPES/STM study concerning the
existence of a gap consistent with a generally d-wave angle-dependence.
However, the higher resolution study revealed that this gap has what appears
to be a two-component structure, in that the gap near the anti-nodal point
at $(\pi,0)$, is considerably larger (by a factor of 2-3) than a simple
extrapolation from the nodal direction would suggest. Since there are
several forms of density wave order known to be present in {La$_{2-x}$Ba$_x$%
CuO$_4$}, it is not straightforward to unambiguously identify particular
spectral features with particular types of order. This is particularly
problematic, since fluctuating order can also lead to a pseudo-gap with many
similarities to the gap that would be produced in the corresponding ordered
phase.

That having been said, it is striking how much the observed spectrum in {La$%
_{2-x}$Ba$_x$CuO$_4$} resembles the mean-field spectrum found under the
assumption that there is large amplitude PDW (striped) order coexisting with
small amplitude uniform superconducting order. The PDW has a large gap along
the anti-nodal [($\pi$,0) or (0,$\pi$)] direction, and a Fermi pocket in the
nodal [($\pi$,$\pi$)] direction, whose spectral weight is considerable only
along an open ``Fermi arc'' region which nearly coincides with the bare
Fermi surface. If, in addition, there is a uniform d-wave component to the
order parameter, this completes the gapping of the Fermi surface. A spectrum
that resembles the measured ARPES spectrum in {La$_{2-x}$Ba$_x$CuO$_4$} can
be obtained\cite{bergandchen} under the assumption that the PDW has a gap
magnitude $\Delta_{PDW}=40$meV, and uniform gap magnitude $\Delta_d=8$meV.
The ARPES spectrum is always measured at temperatures well above the bulk
superconducting $T_c=4K$, so to the extent that this identification is
correct, what is being observed is a pseudo-gap. Moreover, as mentioned
above, the gap structure persists to temperatures above $T_{SDW}$, although
at the higher temperatures, the gap in the nodal regions decreases
appreciably. Much more detailed analysis of the energy, momentum, and
temperature dependence of the ARPES spectrum will be necessary to test this
interpretation.

\emph{Relevance to other cuprates:} We have summarized the evidence for a
state with the symmetries of a PDW state (or the glassy version of it) in $%
\frac{1}{8}$ doped {La$_{2-x}$Ba$_x$CuO$_4$}. There is also evidence for
dynamical layer decoupling effects in {La$_{1.6-x}$Nd$_{0.4}$Sr$_x$CuO$_{4}$}%
\cite{lnsco}, and also in {La$_{2-x}$Sr$_x$CuO$_4$} in a magnetic field.\cite%
{lake,basov08} (For the possible relevance of these ideas to certain heavy
fermion and organic superconductors see [\onlinecite{foot8}].)

The quasiparticle spectrum of a PDW state has striking qualitative
resemblance to the spectra seen in ARPES in other cuprates, especially BSCCO
2212 and 2201. This possibly sheds new light on the issue of the
``nodal-anti nodal dichotomy'': According to this interpretation, both the
nodal and the anti-nodal gaps are superconducting gaps, the first being
uniform and the other modulated. There are two distinct types of order
(``two gaps''), but they are both superconducting, and so they can smoothly
evolve into one another. Perhaps, as the temperature is decreased, the PDW
gradually decreases and the uniform order parameter increases, while their
sum (which is determined by relatively high energy microscopic physics)
remains approximately constant. (Note that an early study\cite{podolsky} of
modulated structures seen in STM\cite{davis,kapitulnik} concluded that they
could be understood in terms of just such a two-superconducting-gap state.)

More generally, one of the most remarkable features of the pseudo-gap
phenomena is the existence of effects of superconducting fluctuations,
detectable\cite{ong,sri,vadim} for instance in the Nernst signal, over a
surprisingly broad range of temperatures and doping concentrations. At a
broad-brush level\cite{emerykivelson}, these phenomena are a consequence of
a phase stiffness scale that is small compared to the pairing scale.
However, it is generally difficult to understand the existence of such a
broad fluctuational regime on the basis of any sensible microscopic
considerations. The glassy nature of the ordering phenomena in a PDW may
hold the key to this central paradox of HTC phenomenology, as it gives rise
to an intrinsically broad regime in which superconducting correlations
extend over large, but not infinite distances.

\begin{acknowledgments}
We thank John Tranquada, Hong Yao, Ruihua He, Srinivas Raghu, Eun-Ah Kim,
Vadim Oganesyan, Kathryn A. Moler and Shoucheng Zhang for great discussions.
This work was supported in part by the National Science Foundation, under
grants DMR 0758462 (EF) and DMR 0531196 (SAK), and by the Office of Science,
U.S. Department of Energy under Contracts DE-FG02-07ER46453 through the
Frederick Seitz Materials Research Laboratory at the University of Illinois
(EF) and DE-FG02-06ER46287 through the Geballe Laboratory of Advanced
Materials at Stanford University (SAK).
\end{acknowledgments}

\appendix

\section{Order parameters}

\label{app:appendixA}

We will now define the various order parameters introduced in Sec. \ref{op}
and discuss their symmetry properties. The striped superconducting order
parameter $\Delta _{Q}$ is a charge 2e complex scalar field, carrying
momentum $Q$. To define it microscopically, we write the superconducting
order parameter as
\begin{eqnarray}
\Delta \left( \vec{r},\vec{r}^{\,\prime }\right) &\equiv &\left\langle \psi
_{\uparrow }^{\dagger }\left( \vec{r}\right) \psi _{\downarrow }^{\dagger
}\left( \vec{r}^{\,\prime }\right) \right\rangle  \notag \\
&=&F\left( \vec{r}-\vec{r}^{\,\prime }\right) \left[ \Delta _{0}+\Delta
_{Q}e^{i\vec{Q}\cdot \vec{R}}+\Delta _{-Q}e^{-i\vec{Q}\cdot \vec{R}}\right]
\text{,}  \notag \\
&&
\end{eqnarray}%
where $R=\left( \vec{r}+\vec{r}^{\,\prime }\right) /2$, $F\left( \vec{r}-%
\vec{r}^{\,\prime }\right) $ is some short range function (for a ``d-wave
like'' striped superconductor, $F\left( \vec{r}\right) $ changes sign under $%
90^\circ$ rotation), and $\Delta _{0}$ is a uniform order parameter. In the
remaining of this appendix, we set $\Delta_{0}=0$. The effect of $\Delta
_{0} $ is discussed in Appendix \ref{app:uniform}. In cases where a $\Delta
_{\bar{Q}}$ is also non-zero (where $\bar{Q}$ is related to $Q$ by a $%
90^\circ$ rotation) analogous terms with $\Delta _{Q}$ replaced by $\Delta _{%
\bar{Q}}$ have to be added.

The order parameters that may couple to $\Delta _{Q}$ and their symmetry
properties are as follows. The nematic order parameter $N$ is a real
pseudo-scalar field; the CDW $\rho _{K}$ with $K=2Q$ is a complex scalar
field; $\vec{S}_{Q}$ is a neutral spin-vector complex field. All these order
parameters are electrically neutral. Under spatial rotation by $\pi /2$, $%
N\rightarrow -N$, $\rho _{K}\rightarrow \rho _{\bar{K}}$, $\vec{S}%
_{Q}\rightarrow \vec{S}_{\bar{K}}$, and $\Delta _{Q}\rightarrow \pm \Delta _{%
\bar{Q}}$, where $\pm $ refers to a d-wave or s-wave version of the striped
superconductor. Under spatial translation by $r$, $N\rightarrow N$, $%
\rho_{K}\rightarrow e^{iK\cdot r}\rho _{K}$, $\vec{S}_{Q}\rightarrow
e^{iQ\cdot r}\vec{S}_{K}$, and $\Delta _{Q}\rightarrow e^{iq\cdot r}\Delta
_{Q}$.

With these considerations, we write all the possible fourth order terms
consistent with symmetry:
\begin{widetext}
\begin{eqnarray}
\mathcal{F}_{4} =&&+u\left(\vec{S}_{Q}\cdot \vec{S}_{Q}\Delta _{Q}^{\star }\Delta
_{-Q}\vec{+}S_{\bar{Q}}\cdot \vec{S}_{\bar{Q}}\Delta _{\bar{Q}}^{\star
}\Delta _{-\bar{Q}}+\mathrm{c.c.}\right) \notag \\
&&+\left(v_{+}[\vec{S}_{Q}^{\star }\cdot \vec{S}_{Q}+\vec{S}_{\bar{Q}}^{\star }\cdot \vec{S}_{\bar{Q}}]
+\tilde{v}_{+}[|\rho _{K}|^{2}+|\rho _{\bar{K}}|^{2}]\right)
\left( |\Delta _{Q}|^{2}+|\Delta _{-Q}|^{2}+|\Delta _{{\bar{Q}}}|^{2}+|\Delta _{-{\bar{Q}}}|^{2}\right\}  \notag \\
&&+\left( v_{-}[\vec{S}_{Q}^{\star }\cdot \vec{S}_{Q}-\vec{S}_{\bar{Q}}^{\star }\cdot \vec{S}_{\bar{Q}}]
+\tilde{v}_{-}[|\rho _{K}|^{2}-|\rho _{\bar{K}}|^{2}]\right)
 \left( |\Delta _{Q}|^{2}+|\Delta _{-Q}|^{2}-|\Delta _{{\bar{Q}}}|^{2}-|\Delta _{-{\bar{Q}}}|^{2}\right)  \notag \\
&&+vN^{2}\left\{ \left(|\Delta _{Q}|^{2}+|\Delta _{-Q}|^{2}\right)+\left(|\Delta _{\bar{Q}}|^{2}+|\Delta _{-\bar{Q}}|^{2}\right)\right\}  \notag \\
&&+\lambda _{+}\left\{ \left(|\Delta _{Q}|^{2}+|\Delta _{-Q}|^{2}\right)^{2}+\left(|\Delta _{\bar{Q}}|^{2}+|\Delta _{-\bar{Q}}|^{2}\right)^{2}\right\}
+\lambda _{-}\left\{ \left(|\Delta _{Q}|^{2}-|\Delta _{-Q}|^{2}\right)^{2}+\left(|\Delta _{\bar{Q}}|^{2}-|\Delta _{-\bar{Q}}|^{2}\right)^{2}\right\}
+\ldots
\notag \\
&&
\end{eqnarray}%
\end{widetext}
where we have explicitly shown all the terms involving $\Delta _{Q}$, while
the terms $\ldots $ represent the remaining quartic terms all of which, with
the exception of those involving $N$, are exhibited explicitly in Ref. %
\onlinecite{zke}.

On physical grounds, we have some information concerning the sign of various
terms in $\mathcal{F}_{4}$. The term proportional to $u$ determines the
relative phase of the spin and superconducting stripe order - we believe $%
u>0 $ which thus favors a $\pi /2$ phase shift between the SDW and the
striped superconducting order, i.e. the peak of the superconducting order
occurs where the spin order passes through zero. The other interesting thing
about this term is that it implies an effective cooperativity between spin
and striped superconducting order. The net effect, i.e. whether spin and
striped superconducting order cooperate or fight, is determined by the sign
of $|u|-v_{+}-v_{-}$, such that they cooperate if $|u|-v_{+}-v_{-}>0$ and
oppose each other if $|u|-v_{+}-v_{-}<0$. It is an interesting possibility
that spin order and superconducting stripe order can actually favor each
other even with all `` repulsive" interactions. The term proportional to $%
\lambda _{-}$ determines whether the superconducting stripe order tends to
be real ($\lambda _{-}>0$), with a superconducting order that simply changes
sign as a function of position, or a complex spiral, which supports
ground-state currents ($\lambda _{-}<0$).

\section{Coexisting uniform and striped order parameters}

\label{app:uniform}

In this appendix, we analyze the coupling of a striped superconducting order
parameter $\Delta _{Q}$ to a uniform order parameter, $\Delta _{0}$. In this
case, we have to consider in addition to the order parameters introduced in
Sec. \ref{op} a CDW order parameter with wavevector $Q$, denoted by $\rho
_{Q}$. The additional cubic terms in the Ginzburg-Landau free energy are
\begin{widetext}
\begin{eqnarray}
\mathcal{F}_{3,u}=
\gamma _{Q}\Delta _{0}^{\star }\left[ \rho _{Q}\Delta
_{-Q}+\rho _{Q}^{\star }\Delta _{Q}+\rho _{\bar{Q}}\Delta _{-\bar{Q}}
+\rho _{\bar{Q}}^{\star }\Delta _{\bar{Q}}\right] +\mathrm{c.c.}
+g_{\rho }\left[ \rho _{2Q}^{\star }\rho _{Q}^{2}+\rho _{2\bar{Q}}^{\star}\rho _{\bar{Q}}^{2}+\mathrm{c.c.}\right] \text{.}
\label{Fu3}
\end{eqnarray}%
Eq. \eqref{Fu3} shows that if both $\Delta _{0}$ and $\Delta _{Q}$ are
non-zero, there is necessarily a coexisting non-zero $\rho _{Q}$, through
the $\gamma _{Q}$ term. The additional quartic terms involving $\Delta _{0}$
are
\begin{eqnarray}
\mathcal{F}_{4,u} &=&u_{\Delta }\left( \Delta _{0}^{\star 2}\Delta_{Q}\Delta _{-Q}
+\Delta _{0}^{\star 2}\Delta _{\bar{Q}}\Delta _{-\bar{Q}}+\mathrm{c.c.}\right)
+\delta |\Delta _{0}|^{2}[|\Delta _{Q}|^{2}+|\Delta _{\bar{Q}}|^{2}]
\notag \\
&+&|\Delta _{0}|^{2}\left[ u_{\rho }
\left( \left\vert \rho _{Q}\right\vert^{2}+\left\vert \rho _{\bar{Q}}\right\vert ^{2}\right)
+u_{\rho }^{\prime}\left( \left\vert \rho _{2Q}\right\vert ^{2}+\left\vert \rho _{2\bar{Q}}\right\vert ^{2}\right) \right]
+v^{\prime }|\Delta _{0}|^{2}[\vec{S}_{Q}^{\star }\cdot \vec{S}_{Q}
+\vec{S}_{\bar{Q}}^{\star }\cdot \vec{S}_{\bar{Q}}]\text{.}
\label{Fu4}
\end{eqnarray}%
\end{widetext}
Let us now consider the effect of quenched disorder. Following the
discussion preceding Eq. \eqref{thetapm}, we write the order parameters in
real space as $\Delta \left( r\right) =\left\vert \Delta _{0}\right\vert
e^{i\theta _{0}}+\left\vert \Delta _{Q}\right\vert e^{i\left(
\theta_{Q}+Q\cdot r\right) } +\left\vert \Delta _{-Q}\right\vert e^{i\left(
\theta_{-Q}-Q\cdot r\right) }$ and $\rho \left( r\right) =\left\vert
\rho_{Q}\right\vert \cos \left( Q\cdot r+\phi _{Q}\right) +\left\vert
\rho_{2Q}\right\vert \cos \left( 2Q\cdot r+\phi \right) $. Let us assume
that the disorder nucleates a point defect in the CDW, which in this case
corresponds to a $2\pi $ vortex in the phase $\phi _{Q}$. By the $g_{\rho }$
term in Eq. \eqref{Fu3}, this induces a $4\pi $ vortex in $\phi $. (Note
that a in the presence of $\rho _{Q}$, a $2\pi $ vortex in $\phi $ is not
possible.) The $\gamma _{\Delta }$ term in Eq. \eqref{F3gamma} then dictates
a $2\pi $ vortex in the phase $\theta _{-}=\left( \theta _{Q}-\theta
_{-Q}\right) /2$. However, unlike before, this vortex does not couple to the
global superconducting phase $\theta _{+}=\left( \theta _{Q}+\theta
_{-Q}\right) /2$. Therefore, an arbitrarily small uniform superconducting
component is sufficient to remove the sensitivity of a striped
superconductor to disorder, and the system is expected to behave more or
less like a regular (uniform) superconductor.



\begin{thebibliography}{99}
\bibitem{foot1} The term ``pair-density-wave'' has been used in the
literature to describe two different states. In Ref. [\onlinecite{PDW}] it
was used to describe a state with coexisting superconducting and CDW order.
While there may be limits where this term conveys a useful intuitive picture
of the microscopic nature of the state, it is not a truly distinct state
since it has a non-vanishing uniform component of the SC order parameter and
the period of the modulations of the SC order is the same as that of the
CDW. The PDW state described here is the same as that of Ref. [%
\onlinecite{agterberg08}], namely a state in which the superconducting order
parameter is modulated (such that its average vanishes exactly).
\bibitem{foot2} Close in spirit to the present work is the staggered
superconducting state found in a study of the 1D Kondo lattice\cite{zachar},
and in various studies of the superconducting state with odd-frequency
pairing\cite{oddstaggered}. There have also been some interesting papers on
a somewhat different, but very attractive mechanism of PDW formation in
systems with multiple Fermi surfaces and an appropriate nesting condition%
\cite{Erica,othermultipleFSreferences}. Recent work found evidence for
coexistence of superconducting correlations and incommensurate charge order
in the 1D Hubbard model with bond-charge interaction.\cite{aligia07}
\bibitem{foot3} The SO(5) spiral generalizes the notion of an
antiferromagnetic (SO(3)) spiral, and for the same reasons\cite{pryadko} is
relatively difficult to realize in any microscopic model in the absence of
spin-orbit coupling.
\bibitem{foot4} One can also imagine similar interplay between between PDW
order and more exotic states, such as DDW\cite{chakravarty}, which follows
similar lines as the interplay with SDW order.
\bibitem{foot7} When $\rho \Delta \sim 1$, the basis for the various
expressions for $J_{n}$ are violated, as it is then no longer well justified
to treat the density of states as a constant. However, this should not
change the qualitative conclusions.
\bibitem{foot8} Recent experiments suggest that an FFLO state, a close
relative of the PDW, may be present in the CeCoIn$_5$ heavy fermion
superconductor\cite{CeCoIn5,kenzelmann-science} and also in $\kappa$%
-(BEDT-TTF)$_2$Cu(NCS)$_2$ organic superconductors\cite{kappa-BEDT-SC}. The
phenomenology of the PDW discussed here, particularly the sensitivity to
disorder, should then apply to these systems as well.
\end{thebibliography}
\end{document}